\documentclass[12pt]{article}
\usepackage{graphicx,psfrag,epsf}
\usepackage{enumerate}
\usepackage{natbib}
\usepackage{url} 
\usepackage{xcolor} 
\usepackage{amsthm,amsmath}
\usepackage{amssymb,amsfonts}
\usepackage{graphicx,psfrag,epsf}
\usepackage{booktabs}
\usepackage{mdframed}
\usepackage{glossaries}
\RequirePackage{bbm}
\RequirePackage{booktabs,array,hhline,bm,tcolorbox,mathtools}
\usepackage{placeins,float}
\usepackage{comment}
\usepackage{multirow}
\usepackage{lineno}

\usepackage[colorlinks=true, linkcolor=blue,citecolor=cyan]{hyperref}

\definecolor{xwcommentcolor}{RGB}{244, 143, 177} 

\newcommand{\blind}{0}

\begin{document}
\nolinenumbers

\def\spacingset#1{\renewcommand{\baselinestretch}%
{#1}\small\normalsize} \spacingset{1}

\def\B0{\mbox{\boldmath $0$}}
\def\Ba{\mbox{\boldmath $A$}}
\def\Bb{\mbox{\boldmath $B$}}
\def\Bc{\mbox{\boldmath $C$}}
\def\Bd{\mbox{\boldmath $D$}}
\def\Bl{\mbox{\boldmath $L$}}
\def\Bn{\mbox{\boldmath $N$}}
\def\Br{\mbox{\boldmath $R$}}
\def\Bt{\mbox{\boldmath $T$}}
\def\Bv{\mbox{\boldmath $V$}}
\def\Bx{\mbox{\boldmath $X$}}
\def\By{\mbox{\boldmath $Y$}}
\def\Bz{\mbox{\boldmath $Z$}}

\def\btheta{\mbox{\boldmath $\theta$}}
\def\htheta{\mbox{$\hat{\theta}$}}

\def\balpha{\mbox{\boldmath $\alpha$}}
\def\bbeta{\mbox{\boldmath $\beta$}}
\def\bdelta{\mbox{\boldmath $\delta$}}
\def\bepsilon{\mbox{\boldmath $\epsilon$}}
\def\bGamma{\mbox{\boldmath $\Gamma$}}
\def\blambda{\mbox{\boldmath $\lambda$}}
\def\bLambda{\mbox{\boldmath $\Lambda$}}
\def\bmu{\mbox{\boldmath $\mu$}}
\def\bSigma{\mbox{\boldmath $\Sigma$}}
\def\btau{\mbox{\boldmath $\tau$}}
\def\btheta{\mbox{\boldmath $\theta$}}

\definecolor{steelblue}{RGB}{70, 130, 180}
\definecolor{darkorange}{RGB}{255, 140, 0}
\definecolor{salmon}{RGB}{249, 120, 110}


\if0\blind
{
  \title{\bf From Average Effects to Proportions Benefiting: A Tutorial on ETZ-Powered Counterfactual Individual Benefit Analysis}
   \author{Xingya Wang\thanks{
    Corresponding author: xw495@cam.ac.uk
  }\\ {\small https://orcid.org/0000-0003-1532-9952}, {\small xw495@cam.ac.uk}
  \\
    {\small MRC Biostatistics Unit, University of Cambridge}\\
    {\small Department of Mathematics, The University of Manchester}\\
    Yang Han\\ {\small https://orcid.org/0000-0002-4143-7765}, 
    {\small yang.han@manchester.ac.uk}\\
    {\small Department of Mathematics, The University of Manchester}\\
     Yushi Liu\\
     {\small liu\_yushi@lilly.com}\\
    {\small Eli Lilly and Company,}\\
    Szu-Yu Tang\\
    {\small Szu-Yu.Tang@pfizer.com}\\
    {\small Pfizer Worldwide Research \& Development}\\
   Jason C. Hsu\thanks{
    Corresponding author: Hsu.1@osu.edu
  }\\ {\small https://orcid.org/0000-0002-1277-0403}, 
   {\small Hsu.1@osu.edu}\\
   {\small Department of Statistics, The Ohio State University}\\
   }
 
  \maketitle
} \fi

\if1\blind
{
  \bigskip
  \bigskip
  \bigskip
  \begin{center}
    {\LARGE\bf From Average Effects to Proportions Benefiting: A Tutorial on ETZ-Powered Counterfactual Individual Benefit Analysis}
\end{center}
  \medskip
} \fi

\bigskip
\begin{abstract}

Consider the question: ``Is it true that at least 80\% of patients would benefit more from receiving a new treatment, $Rx$, than from receiving a standard treatment, $C$?'' Conventional treatment efficacy analyses based on average treatment effects cannot directly answer this question. 
Yet, even if the true mean treatment effects were known exactly, the answer to this question could still be either “yes” or “no”, depending on the variability in patients' treatment responses and the dependence between their potential outcomes.

This tutorial introduces a proportion-based perspective on individualized treatment benefit through the ETZ-powered Counterfactual Individual Benefit Analysis framework. We present two metrics, the individual benefit probability and the individual benefit lower bound, which enable confidence-based assessments of whether a prespecified proportion of patients achieves a clinically meaningful treatment benefit. We develop these metrics in both the factual random-pair and counterfactual same-patient settings. In particular, the ETZ decomposition provides a transparent characterization of different sources of outcome variability and enables more precise uncertainty quantification for counterfactual treatment comparisons. We illustrate the framework and its practical implementation using patient-level data from a Phase II randomized clinical trial.

By shifting the focus from average treatment effects to population proportions of individual benefit, this tutorial provides a complementary perspective for evaluating treatment efficacy and informing individualized treatment decisions while controlling the type I error rate at a prespecified level.
\end{abstract}

\noindent%

\section*{Conflicts of interest}

X.W., Y.H., S-Y.T., J.C.H. declare no competing interests.\\
Y. L. is an employee and shareholder of Eli Lilly.

\section*{Funding information}
This research received no specific funding from any funding agency in the public, commercial, or not-for-profit sectors.

\vfill

\newpage
\spacingset{1.45} 

\section{Motivation and background}
To personalize medicine, ideally, we should know what causes a person to be sick.  
If a \emph{plausible mechanism} is known, then lines 117-121 of \cite{FDA2026Individualized} state that a pathway to developing a drug and gaining marketing approval is to prove the therapy causes a 
\begin{quote}
	``change in disease \emph{trajectory} that is inconsistent with the natural history of the disease $\ldots$ (with) improvement in outcome (that) cannot reasonably be attributed to alternative treatments or natural variability in the disease phenotype''.
\end{quote}

\subsection{A taxonomy of individualized medicine}
The term \emph{individualized medicine} is used in the literature to describe a broad range of medical and statistical strategies. 
These strategies differ in the degree to which treatment is individualized: some therapies are manufactured for a single patient, some are designed for subgroups defined by causal mutations, some target patients expressing a biological marker, and some assign treatments according to observed covariates. 
To clarify the scope of this article, Table~\ref{tab:iMedicineTaxonomy} summarizes these different forms of individualized medicine, ordered from the highest to the lowest degree of individualization, and highlights where ETZ modeling contributes to the corresponding statistical problem.
\begin{table}[htbp]
	\resizebox{\textwidth}{!}{
	\begin{tabular}{|c|c|c|c|c|}
	\hline
	\textbf{iMedicine name} & \textbf{Treating} & for \textbf{Example} & for \textbf{Treating} & \textbf{ETZ's advantage} \\
	\hline 
	\textbf{Bespoke therapies} & Individual patient & CAR-T & Multiple Myeloma &  \\
	\hline
	\textbf{Individualized therapies} & Subgroups defined by causal mutation & EXONDYS51\texttrademark & Duchenne Muscular Dystrophy & Counterfactual UQ \\
	\hline
    \textbf{Targeted therapie}s & Patients expressing drug target & Anti-amyloid-$\beta$ monoclonal antibodies & Alzheimer's disease & Confident prediction \\
	\hline
	\textbf{Individual Treatment Rule} & Subgroups defined by covariates & Any therapy & Any disease &  \\
	\hline
\end{tabular}
}
	\caption[Classifying medicine by degree of individualization]{Classifying medicine by degree of individualization: in \textit{descending} order of individualization from top to bottom, and from left to right examples of medical treatments, the diseases they treat, and key contributions enabled by the ETZ modeling of this article.}
    \label{tab:iMedicineTaxonomy}
\end{table}

\paragraph{Bespoke therapies treating each individual patient}


Chimeric Antigen Receptor T‑cell (CAR‑T) therapy for multiple myeloma is a personalized immunotherapy in which each patient’s own T‑cells are collected, genetically engineered to recognize the BCMA protein on myeloma cells, expanded in a lab, and then reinfused back to that patient to attack their (hematologic, non‑solid tumor) cancer. 

\paragraph{Individualized therapies treating causal mutations}


Duchenne muscular dystrophy (DMD) is a severe, progressive muscular disease caused by mutations in the dystrophin gene.  
DMD mutations often delete one or more exons in the dystrophin gene, which breaks the reading frame. When the frame is broken, the cell produces no usable dystrophin, and muscle fibers progressively degenerate.

Antisense oligonucleotides (ASOs) such as those in Table \ref{tab:ASOs} target specific genetic mutations that cause a disease are examples of what the FDA calls \textit{Individualized Therapies}. 
\begin{table}[tbh]
	\centering
	\begin{tabular}{lcc}
		\hline
		\textbf{Drug} & \textbf{Exon Target} & \textbf{Indicated for patients with} \\
		\hline
		Exondys~51 (eteplirsen) & Exon 51 & mutations amenable to exon 51 skipping \\
		Vyondys~53 (golodirsen) & Exon 53 & mutations amenable to exon 53 skipping \\
		Amondys~45 (casimersen) & Exon 45 & mutations amenable to exon 45 skipping \\
		\hline
	\end{tabular}
	\caption{Summary of Sarepta's FDA-approved exon-skipping therapies for DMD}
	\label{tab:ASOs}
\end{table}

Mutation‑specific, not patient‑specific, they are designed for subgroups of patients who share the same 
mutation (e.g., those who need exon 45, 51, or 53 of the dystrophin gene skipped). 
Exon‑skipping drugs for DMD that are ASOs work by 
\begin{enumerate}
	\item masking that exon so the splicing machinery skips it, restoring the reading frame,
	\item thus enabling production of a shorter but functional dystrophin protein, 
	\item and the increased dystrophin level helps slow muscle fiber damage.
\end{enumerate}
Even a small increase in dystrophin slows degeneration and inflammation that drives DMD.  
Each therapy does not cure the disease, but it shifts the biology toward a milder phenotype.

Lines 120-121 of \cite{FDA2026Individualized} states 
\begin{quote}
	\textit{data submitted by the sponsor must adequately demonstrate that the improvement in outcome cannot reasonably be attributed to alternative treatments or natural variability in the disease phenotype}
\end{quote}
This requirement in proving treatment efficacy is challenging for DMD because of 
high variability in clinical outcome measures (such as the \textit{6‑minute walk distance} and \textit{time to climb 4 stairs}), and the relatively small sample size available for testing treatment effects.  
A rural patient may be more fatigued traveling to the hospital than another living in an urban area, a boy may have had a good breakfast while another may not have, as explained by Dr. Mark Levenson (FDA/CDER/OB) in his keynote address at the 2026 Duke Industry Statistics Symposium. 

Fortunately, our ETZ modeling principle helps to overcome the challenge.  
While traditional MMRM (mixed model repeated measures) statistical analysis separates total outcome uncertainty to \emph{between} and \emph{within} patient variability, 
ETZ further separates within patient outcome uncertainty into \emph{before-treatment} patient variability and \emph{after-treatment} disease trajectory variability.  
ETZ modeling not only quantifies the UQ reduction achieved by assessing treatment effect using change-from-baseline, but also further reduces UQ by making each patient their intrinsic twin.  

\paragraph{Targeted therapies}
In the words of FDA/CDER ex-director Janet Woodcock, \textit{targeted therapies} are personalized/precision medicine.  
Such medicine includes large molecule monoclonal antibodies (compounds with a {\em mab} suffix) targeting specific antigens on cell surfaces, 
e.g., atezolizu\textit{mab} blocks program death ligand PDL-1 on solid tumor cancer cell surfaces, or small molecules penetrating cell membranes to interact with enzymatic activities such as tyrosine kinase (compounds with a {\em tinib} suffix), e.g., ibru\textit{tinib} inhibits Bruton's tyrosine kinase in patients with chronic lymphocytic leukemia (CLL).
As medicine enters an era of precision, the challenge for Statistics is no longer whether personalized medicine is possible, but how best to translate its potential into clinical practice.

Targeted therapies tend to naturally have patient subgroups, divided according to the extent patients have the intended drug target expressed or presented.  
The main challenge of developing targeted therapies is how best to choose the patient subgroup to target.
Developing a targeted therapy typically involves the co-development of a Companion Diagnostic test (\(CD_x\)), dividing patients into a marker-positive ($g^+$) subgroup and its complementary marker-negative ($g^-$) subgroup, corresponding to patients who are more likely to benefit, and those who are less likely to.  
\subsection{Existing statistical approaches to personalized medicine}
\subsubsection{Individual Treatment Rule (ITR)}
\cite{zhao2015new} formulated the personalized medicine problem as finding the optimal individual treatment rule (ITR) by maximizing the expected clinical responses.
More recently, \cite{lei2021conformal} developed interval estimates for individual treatment effects using conformal inference in observational studies.
There is also a line of research on Digital Health using data from wearable devices.

Without considering variance, it is impossible to incorporate uncertainty or to address the probability-based questions of interest. 
The optimal individualized treatment rule (ITR) proposed by \citet{zhao2015new} is a representative \emph{expectation-based} method for efficacy comparison, which focuses solely on maximizing the mean outcome and does not account for the variabilities in outcomes.
In Section \ref{sec:CounterexampleITR}, we provide two counterexamples to illustrate this point in more detail.

The general goal of the \emph{expectation-based} method is to identify the treatment rule that maximizes the \emph{expected clinical outcome}. 
Let $X$ denote patient covariates, $\mathrm{Trt} \in \{Rx,C\}$ the treatment assignment, and $Y$ a nonnegative clinical outcome (shifted so that higher values indicate better responses). 
For any decision rule $d(X)$ which is a mapping that assigns a treatment $\mathrm{Trt}$ based on $X$, outcome weighted learning (OWL) seeks the optimal rule $d^{\ast}(X)$ that maximizes the expected outcome $\mathbb{E}[Y(d)\,|\,X]$, that is,
\[
d^{\ast}(X) = \arg\max_{d} \, \mathbb{E}[\,Y(d)\,|\,X\,], 
\qquad d^{\ast}: X \rightarrow \{Rx,C\}.
\]

\subsubsection{Conformal prediction}
\cite{lei2021conformal} approached the personalized medicine problem from a different perspective.
Although they have not proposed a specific treatment assignment rule, they provided conformal prediction intervals for the counterfactual outcome or the individual treatment effect of the next patient as a decision-support tool.
The concept of conformal prediction was proposed to quantify how differently the next observation performs compared to the previous trend in online testing.(\cite{shafer2008tutorial})
The features of online testing are summarized in Appendix \ref{app:OnlineTesting}.
They use a non-conformity score designed to capture the impact of including the next observation in the ``bag'' of previous observations on the quantity of interest.
A low non-conformity score indicates that the candidate value of the next observation behaves ``conformally'' relative to the historical observations.

In the simplest example of how one calculates a Conformity score, whether \(z_{20}\) conforms to \(z_1\), … , \(z_{19}\), one would say \(z_{20}\) “conforms” to \(z_1\), … , \(z_{19}\) if \(z_{20}\) is not far from the average of all \(z\), and within the maximum deviations (from average) of \(z_1\), … , \(z_{19}\). 

The two methods discussed above, the individual treatment rule (ITR) and the weighted conformal approach, and our proposed method focus on different population-level targets and therefore provide inferences for different aspects of personalized medicine decision-making.
The ITR approach targets the expectation, as its treatment assignment strategy is designed to maximize patients’ expected outcomes.
In contrast, the weighted conformal approach focuses on inference for the next observation, reflecting the origins of conformal prediction in online testing settings where prediction for a future individual is of primary interest.
Our method instead quantifies the individual benefit probability, which provides inference on the proportion of the population that can truly benefit from receiving $ Rx $ instead of receiving $ C $.
For an individual patient, we quantify the probability that they benefit more from the treatment of interest, at a specified confidence level.

\subsubsection{Personalized medicine with wearable devices}
There is also a line of personalized medicine that detects the individual treatment effect using wearable devices.

Personalization often meant stratifying patients by genotype, biomarker status, or clinical risk score. 
The newer digital-health literature broadens that model by using wearable sensors, smartphones, remote monitoring, and computational analytics to characterize each person’s physiology and behavior longitudinally and in real-world settings.
For example, \cite{tyler2020realtime} focuses on making medicine more real-time and more individualized through continuous wearable sensing.

Since the line of personalized medicine using wearable devices is rather distant from the setting of our research, we do not discuss it in detail.

\subsubsection{Problem formulation for Personalized Medicine in this article}\label{sec:ProblemFormulation}
Different from the above lines of research, this article contributes to personalized medicine by addressing uncertainty quantification in randomized controlled trials (RCTs) and in both factual and counterfactual settings.
Potential-outcomes approaches provide a natural counterfactual framework for formulating causal questions in pharmacometrics. 
For example, if $Y_i(Trt)$ denotes the potential outcome for patient $i$ under intervention $t$, population-level quantities such
as
\[
P\{Y_i(t)<q\},
\]
where $q$ is a prespecified patient outcome threshold, characterize the proportion of patients who would attain a specified outcome level under that intervention
\cite{rogers2023causal}. Such quantities characterize the outcome distribution
under one intervention at a time.

Our interest is instead explicitly comparative. Consider a randomized controlled
trial with two treatment arms, where $Rx$ denotes the treatment of interest and
$C$ denotes the control. For patient $i$, let $Y_i(Rx)$ and $Y_i(C)$ denote the
potential outcomes under $Rx$ and $C$, respectively. In the counterfactual
setting, define the individual treatment contrast
\[
D_i^{cf}
:=
Y_i(Rx)-Y_i(C).
\]
For a prespecified clinically meaningful difference margin $\delta$, our primary
population-level question is whether
\[
P\left(D_i^{cf}>\delta\right)\geq\beta,
\]
where $\beta\in(0,1)$ is a prespecified target proportion. For example, taking
$\beta=0.8$, the question becomes: is it true that at least 80\% of patients
would benefit more from receiving $Rx$ than from receiving $C$ by more than
$\delta$?

The remainder of this article is organized as follows.
Section~\ref{sec:CounterexampleITR} examines the limitations of expectation-based individualized treatment rules.
Section~\ref{sec:IBAnalysis} develops the proposed individual benefit framework, first in the factual setting in Section~\ref{sec:IBMethodFactual}, then in the counterfactual setting in Section~\ref{subsec:counterfactual_individual_benefit}, followed by a data example in Section~\ref{sec:IBExample}.
Section~\ref{sec:discussion_future_work} concludes with discussion and directions for future work.

\section{Counterexamples for ITR in three depths}\label{sec:CounterexampleITR}
For a given patient, the same question of interest can be rephrased as:  
``Is it true that the probability of benefiting more from receiving $Rx$ than from receiving $C$ by at least an amount $\delta$ is greater than 80\%?''

The counterfactual framework proposed by \cite{rubin1978bayesian} is introduced in this article because patients want to know what would happen when they took both $ Rx $ and $ C $ but they can only receive one treatment ($ Rx $ or $ C $) factually.
In the counterfactual framework, let $ Trt_i $ be the random variable that assigns the $ i^{th} $ patient to either $ Rx $ or $ C $.
The patient $ i $ has two potential outcomes $ Y_i (Rx) $ and $ Y_i (C) $.

The treatment assignment of patients is ignorable in the counterfactual framework.
Let $ X $ represent relevant covariates in general.
For each given $ X = x $, the random assignment variable $ Trt_i $ is \emph{independent} of the \textit{potential} outcomes $ Y_i (Rx) $ and $ Y_i (C) $.

ETZ-powered counterfactual framework proposed by \cite{wang2025counterfactual} achieves counterfactual uncertainty quantification through factual measurements.

\subsection{Three depths of treatment efficacy comparison}
Bioequivalence was the first area to recognize true ``individualization''.
For a long time, nearly everything outside bioequivalence has been assessed at the average level, and this is still largely the case today.

Inspired by the framework of bioequivalence, we recognize that treatment efficacy comparison can also be understood at three depths: average depth, population depth, and individual depth.
Research in personalized medicine should be able to show that the treatment of interest $Rx$ is superior to the control treatment $C$ in the individual depth.
In this article, we aim to quantify the probability that the treatment of interest $Rx$ is superior to $C$ in the individual depth.

Bioequivalence seeks to answer the question: ``Are these two drugs really the same?''
\cite{fda2001bioequivalence} provides careful definitions of average bioequivalence, population bioequivalence, and individual bioequivalence.
Intuitively, average bioequivalence means that the generic drug and the brand-name drug have sufficiently similar average responses when considered across the study population as a whole.
The population bioequivalence asks whether the two drugs are similar not only in their average response but also in the overall variability of the response in the population.
Individual bioequivalence is the strongest notion, as it asks whether the two drugs are sufficiently similar for the same individual patient, so that switching from one product to the other would not significantly change that patient's response.

Analogously, efficacy analysis is also in three depths.
Average efficacy concerns whether the treatment of interest improves the mean outcome relative to the control when one evaluates the population as a whole.
Population efficacy moves beyond the mean and considers whether the treatment demonstrates an overall advantage at the population level, taking into account how responses are distributed across patients.
Individual efficacy is the deepest notion: it concerns whether, for a given patient, the treatment of interest would lead to a better outcome than the control, thereby capturing treatment benefit at the personalized level.

\subsection{The setting for individualized treatment rules (ITR)}

An individualized treatment rule (ITR) is a map from the covariate space to the treatment space. Following \cite{zhao2015new}, the optimal individualized treatment rule (OITR) assigns to each patient the treatment that maximizes the expected clinical outcome.

Under the counterfactual framework, consider a two-arm randomized controlled trial with treatments $C$ (control) and $Rx$ (active treatment). Let $X_i$ denote the covariates for patient $i$, let $Trt_i \in \{C, Rx\}$ denote the treatment actually received, and let
\[
Y_i(Rx) \quad \text{and} \quad Y_i(C)
\]
denote the potential outcomes under $Rx$ and $C$, respectively. Under randomization,
\[
Trt_i \perp \{Y_i(Rx), Y_i(C)\} \mid X_i.
\]
The OITR is defined by
\[
d^{\ast}(X)
= \arg\max_{t \in \{C,Rx\}} E\!\left\{Y_i(t)\mid X\right\}.
\]
Thus, the OITR depends only on conditional mean outcomes. In the special case with no covariates, the OITR reduces to assigning every patient the treatment with the larger marginal mean outcome.

For the counterexamples below, define the individual treatment effect
\[
\Delta_i = Y_i(Rx)-Y_i(C),
\]
and let the proportion of patients who benefit from $Rx$ by at least a clinically meaningful threshold $\delta$ be
\[
P(\Delta_i>\delta).
\]
For illustration purpose, we take $\delta=0$. 
The OITR cannot guarantee optimal treatment decisions for a substantial proportion of patients under either average depth or population depth. In Section \ref{sec:CounterexamplePop}, we show that the OITR cannot provide such a guarantee even in a favorable population-depth setting, where the two treatment arms are assumed to have equal variances. A corresponding counterexample for average depth is provided in Appendix \ref{sec:CounterexampleAvg}.





\subsection{A counterexample for individualized treatment rules at the population depth}\label{sec:CounterexamplePop}
Even under the favorable scenario where the variances across treatment arms are equal, the OITR cannot guarantee that the treatment decision made for an individual patient is correct.
Here we show a counterexample for population depth.

Now consider the two cases in Table \ref{tab:twocases_percentage}. 
In both cases, $E\!\left[Y_i(Rx)\right] = 2,
E\!\left[Y_i(C)\right] = 1$ and $E[\Delta_i]=1>0.$ Thus, the OITR makes the same decision at the average depth in both cases, namely to assign $Rx$ to every patient.

\begin{table}[htbp]
\centering
\begin{tabular}{|l|c|c|c|c|c|c|}
\hline
Case & meanC & meanRx & sdC & sdRx & $\delta$ & percentage \\ \hline
1 & 1 & 2 & 3 & 3 & 0 & 59.32\% \\ \hline
2 & 1 & 2 & 1.5 & 1.5 & 0 & 68.13\% \\ \hline
\end{tabular}
\caption{Two cases with the same average depth but different population-depth percentages of benefit. Here, percentage denotes $P\{Y_i(Rx)-Y_i(C)>\delta\}$.}
\label{tab:twocases_percentage}
\end{table}

In Case 1, assume
\[
Y_i(Rx)\sim N(2,9),
\qquad
Y_i(C)\sim N(1,9),
\]
with independence between $Y_i(Rx)$ and $Y_i(C)$. Then
\[
\Delta_i = Y_i(Rx)-Y_i(C)\sim N(1,18),
\]
and hence
\[
P(\Delta_i>0)
=
P\!\left(Z>\frac{0-1}{\sqrt{18}}\right)
=
P\!\left(Z>-\frac{1}{\sqrt{18}}\right)
=
\Phi\!\left(\frac{1}{\sqrt{18}}\right)
=
\Phi(0.2357)
\approx 0.5932.
\]

In Case 2, assume instead
\[
Y_i(Rx)\sim N(2,2.25),
\qquad
Y_i(C)\sim N(1,2.25),
\]
again with independence between $Y_i(Rx)$ and $Y_i(C)$. Then
\[
\Delta_i = Y_i(Rx)-Y_i(C)\sim N(1,4.5),
\]
and hence
\[
P(\Delta_i>0)
=
P\!\left(Z>\frac{0-1}{\sqrt{4.5}}\right)
=
P\!\left(Z>-\frac{1}{\sqrt{4.5}}\right)
=
\Phi\!\left(\frac{1}{\sqrt{4.5}}\right)
=
\Phi(0.4714)
\approx 0.6813.
\]

Hence, the two cases are identical at the average depth, since both have
\[
E[\Delta_i]=1>0,
\]
but 0.5932 and 0.6813 differ at the population depth.
Even in the second case, about
\[
1-0.6813 = 0.3187
\]
of patients do not benefit from $Rx$ at the individual level.

\section[Individual benefit analysis]{Individual benefit analysis in ETZ-powered counterfactual framework}\label{sec:IBAnalysis}
\begin{figure}[htbp]
    \centering
\includegraphics[width=0.8\linewidth]{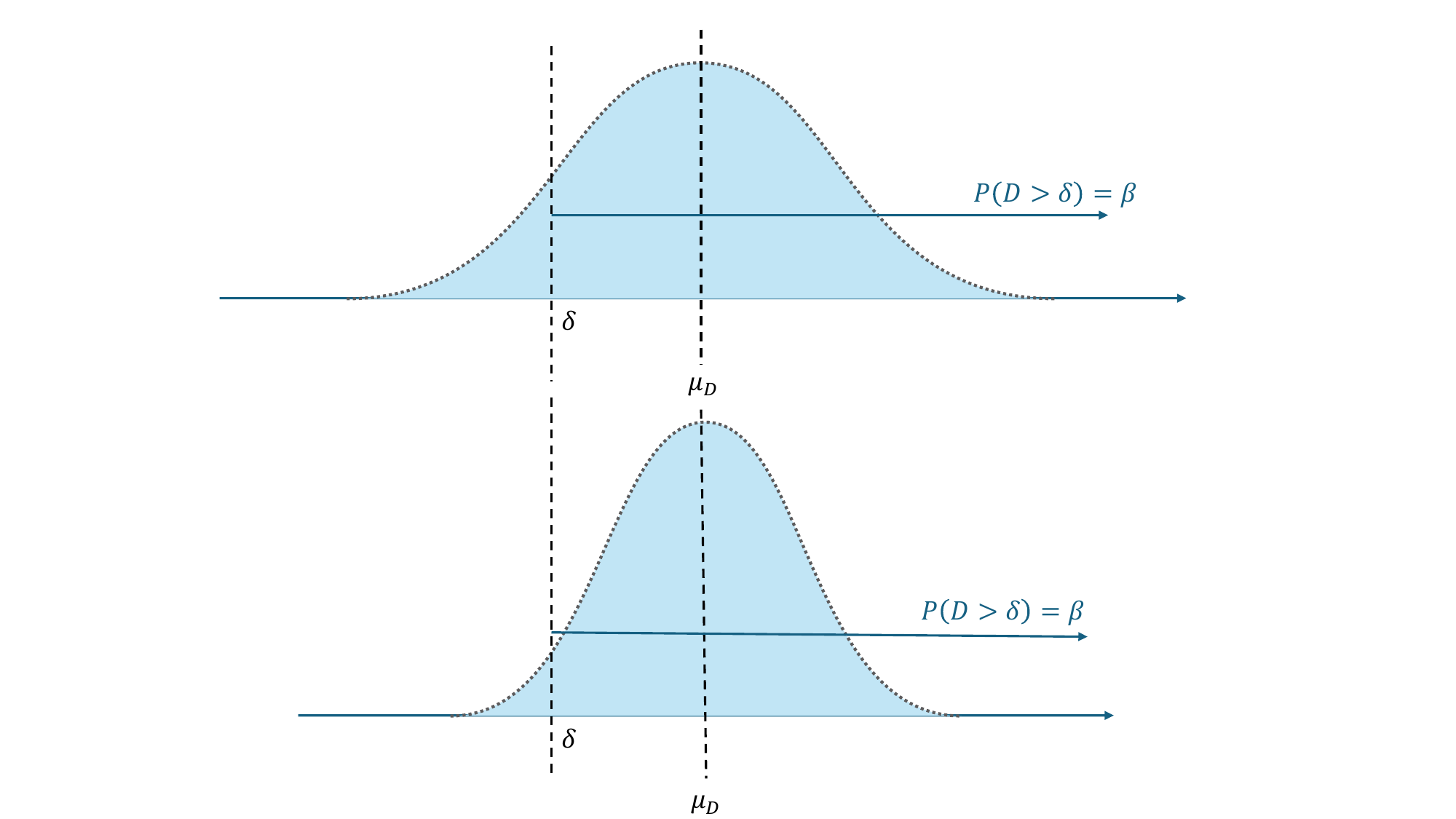}
    \caption[Individualized medicine percentage-perspective problem formulation.]{Plots represent two densities of $D$ in the population space, so areas under the curves to the right of \emph{Win margin} $\delta$ are the \emph{Win probability} $\beta$ of randomly paired patients in which the patient receiving $Rx$ has outcome better than the patient receiving $C$ by at least amount $\delta$.  Even though $\mu_D$ is the same for both top and bottom plots, variability of outcome is higher on top.  Variability matters: conceivably, the answer to the patients' ``Is it true that $\ldots $?'' question may be ``yes'' for the bottom plot, but ``no'' for the top plot.}
    \label{fig:ProblemFormulationPLot}
\end{figure}

We go back to \hyperlink{q:main}{the key question of interest}:
\begin{quote}
\textit{Is it true that at least \(\beta \) proportion of patients will derive a benefit from receiving \(Rx\) rather than \(C\) that exceeds a clinically meaningful margin \(\delta\)?}
\end{quote}
The Figure \ref{fig:ProblemFormulationPLot} illustrates the above key question of interest.

The question can be answered in a mathematical formulation at 95\% confidence level.
In Section \ref{sec:IBMethodFactual}, we introduced the individual benefit analysis and then used the Phase II clinical trial described in \cite{schnell2017subgroup} as an example to illustrate it.
The individual benefit analysis is powered by the ETZ modelling principle because we use it for uncertainty decomposition.

\subsection{Individual benefit probability and individual benefit lower bound in the factual setting}\label{sec:IBMethodFactual}
We first consider the notion of \emph{factual efficacy}. 
In a two-arm treatment setting, factual efficacy is defined as the expected difference in change-from-baseline outcomes between a randomly selected patient from the treatment arm \(Rx\) and an independently selected patient from the control arm \(C\).
Formally, we consider the distribution of the difference between two independent future change-from-baseline outcomes, one drawn from the \(Rx\) arm and the other from the control arm \(C\). Let
\[
\mathcal D_{n_C,n_{Rx}}
=
\Big(
Y^{C[\mathrm{change}]}_1,\ldots,Y^{C[\mathrm{change}]}_{n_C},
Y^{Rx[\mathrm{change}]}_1,\ldots,Y^{Rx[\mathrm{change}]}_{n_{Rx}}
\Big)
\]
denote the observed data, where \(Y^{Trt[\mathrm{change}]}_i\) is the
change-from-baseline outcome for subject \(i\) in treatment arm
\(Trt\in\{C,Rx\}\).

Assume
\[
Y^{C[\mathrm{change}]}_1,\ldots,Y^{C[\mathrm{change}]}_{n_C}
\stackrel{iid}{\sim}
N\!\left(\mu^{C[\mathrm{change}]},(\sigma^{C[\mathrm{change}]})^2\right),
\]
\[
Y^{Rx[\mathrm{change}]}_1,\ldots,Y^{Rx[\mathrm{change}]}_{n_{Rx}}
\stackrel{iid}{\sim}
N\!\left(\mu^{Rx[\mathrm{change}]},(\sigma^{Rx[\mathrm{change}]})^2\right),
\]
and assume independence between the two treatment arms. Let
\(\widetilde Y^{C[\mathrm{change}]}\) and
\(\widetilde Y^{Rx[\mathrm{change}]}\) be independent future draws from
the control and Rx arms, respectively, independent of
\(\mathcal D_{n_C,n_{Rx}}\).
Let factual efficacy contrast \(D\) denote the  difference between the random pair of \(\widetilde Y^{C[\mathrm{change}]}\) and
\(\widetilde Y^{Rx[\mathrm{change}]}\)
\[
D
=
\widetilde Y^{Rx[\mathrm{change}]}
-
\widetilde Y^{C[\mathrm{change}]}.
\]
Factual efficacy is defined as 
\[
\mathbb{E}[D]
=
\mathbb{E}[\widetilde Y^{Rx[\mathrm{change}]}
-
\widetilde Y^{C[\mathrm{change}]}]
= \mu^{Rx[\mathrm{change}]}-\mu^{C[\mathrm{change}]}.
\]
Hence, factual efficacy contrast \(D\) is defined at the individual level while the factual efficacy is defined as the expectation for the overall population.
Then
\begin{equation}
D\sim N\!\left(
\mu^{Rx[\mathrm{change}]}
-\mu^{C[\mathrm{change}]},
\sigma_D^2
\right),\label{eq:Ddist}
\end{equation}
where \(\sigma_D^2\) is the variance of \(D\) and can be decomposed by ETZ modelling principle.

What we call the \textbf{Win Margin} $\delta$ (that is desired by patients and clinicians) in turn defines what we call the \textbf{Win Probability} $\beta = P\{D > \delta\}$ which is a function of the distribution of $D$ in (\ref{eq:Ddist}).  
If $\beta$ is greater than $80\%$, then the answer to our iMedicine question is ``yes'', if not, then the answer is ``no''.  

Let
\[
\theta
=
\big(
\mu^{C[\mathrm{change}]},
\mu^{Rx[\mathrm{change}]},
\sigma^{C[\mathrm{change}]},
\sigma^{Rx[\mathrm{change}]}
\big).
\]

Assuming that the patient with higher outcome is healthier, we defined two quantities: individual benefit probability \(p_{ib}\) and individual benefit lower bound \( L \) to help with individual efficacy justification.
\hyperlink{q:main}{The key question of interest} can be answered through either comparing the individual benefit probability \(p_{ib}\)  with win probability \(\beta\) or comparing the individual benefit lower bound \( L \) with win margin \(\delta\).

For a prespecified Win margin $\delta$, we define the individual benefit
probability $p_{ib}$ as the $100(1-\alpha)\%$ lower confidence bound for the
probability that, for a randomly selected pair of patients, the patient
receiving $Rx$ derives more than $\delta$ additional benefit relative to the
patient receiving $C$. Specifically,
\begin{equation}
p_{ib}
:=
\mathrm{LCB}_{1-\alpha}
\left\{
P_\theta\left(
D > \delta \mid \mathcal{D}_{n_C,n_{Rx}}
\right)
\right\}
=
\mathrm{LCB}_{1-\alpha}
\left\{
\int_{\delta}^{\infty}
f_\theta\left(
d \mid \mathcal{D}_{n_C,n_{Rx}}
\right)\,\mathrm{d}d
\right\},
\end{equation}
where $f_\theta(\cdot \mid \mathcal{D}_{n_C,n_{Rx}})$ denotes the
probability density function of $D$ under $\theta$, conditional on the
observed data $\mathcal{D}_{n_C,n_{Rx}}$, and
$\mathrm{LCB}_{1-\alpha}\{\cdot\}$ denotes the corresponding $100(1-\alpha)\%$ lower confidence bound.

The individual benefit lower bound \(L\) defined here is similar to the idea of clinical meaningful difference, which defines how different can be considered as a sign of efficacy.

For a given win probability \(\beta\),
\begin{equation}
P_\theta\!\left(
D \ge L(\mathcal D_{n_C,n_{Rx}},\beta) \;\middle|\;
\mathcal D_{n_C,n_{Rx}}
\right) \ge \beta
\label{eq:IBPbetarequirement}
\end{equation}
means that, for a randomly selected pair of patients, the probability that the patient receiving \(Rx\) achieves at least \(L\) more benefit than the patient receiving \(C\) is no less than \(\beta\) and here \(L\) depends on both observed data \(\mathcal D_{n_C,n_{Rx}}\) and the required probability \(\beta\).

The requirement in (\ref{eq:IBPbetarequirement}) can only ensure the coverage \(\beta\) conditioned on the observed random sample \(\mathcal D_{n_C,n_{Rx}}\).
If we take the randomness from sampling into consideration and still expect to answer \hyperlink{q:main}{the key question of interest} with a confidence level guarantee, then the requirement becomes
\begin{equation}
\begin{aligned}
& P_\theta\!\left(
P_\theta\!\left(
D \ge L(\mathcal D_{n_C,n_{Rx}},\beta)
\;\middle|\;
\mathcal D_{n_C,n_{Rx}}
\right)
\ge \beta
\right) \\
&\quad =
\int
\mathbf{1}\!\left\{
P_\theta\!\left(
D \ge L(\mathcal D_{n_C,n_{Rx}},\beta)
\;\middle|\;
\mathcal D_{n_C,n_{Rx}}
\right)
\ge \beta
\right\}
\, dP_\theta(\mathcal D_{n_C,n_{Rx}})
\;\ge\;
1-\alpha .
\end{aligned}
\label{eq:IBPbeta-L-requirement-int}
\end{equation}
Throughout this article, the error guarantee refers to a false-positive decision: declaring that the individual benefit requirement is satisfied when it is not. By using a one-sided lower confidence bound, we control the probability of such a decision at level \(\alpha\).

Given $\beta$ and the (unknown) distribution of $D$, solving for $\delta^?$ in $P(D > \delta^?)=\beta$  makes $\delta^?$ a \emph{parameter} so that answer to the question ``$P(D > \delta) > \beta\, ?$'' is ``yes'' if $\delta^? \ge \delta$ , ``no'' otherwise.  
Thus, to control the incorrect decision rate at level-$\alpha$, what we do is to construct a 100$(1-\alpha)\%$ lower confidence bound $\underline{\delta^?}$ 
for $\delta^?$, and answer ``yes'' only when $\underline{\delta^?} > \delta$.   

Given $\delta$ and the (unknown) distribution of $D$, solving for $\beta^?$ in $P(D > \delta)=\beta^?$  makes $\beta^?$ a \emph{parameter} so that answer to the question ``$P(D > \delta) > \beta\, ?$'' is ``yes'' if $\beta^? \ge \beta$, ``no'' otherwise.  
Thus, to control the incorrect decision rate at level-$\alpha$, what we do is to construct a 100$(1-\alpha)\%$ lower confidence bound $\underline{\beta^?}$ 
for $\beta^?$, and answer ``yes'' only when $\underline{\beta^?} \ge \beta$.   

Similarly, the individual benefit probability \(\beta\) which satisfies the requirement, can be computed when the Win Margin \(\delta\) is given.
In this case, the requirement becomes
\begin{equation}
\begin{aligned}
& P_\theta\!\left(
P_\theta\!\left(
D \ge \delta
\;\middle|\;
\mathcal D_{n_C,n_{Rx}}
\right)
\ge p_{ib}(\mathcal D_{n_C,n_{Rx}},\delta)
\right) \;\ge\;
1-\alpha.
\end{aligned}
\label{eq:IBPbeta-L-requirement}
\end{equation}

\begin{figure}[htbp]
    \centering
    \includegraphics[width=0.8\linewidth]{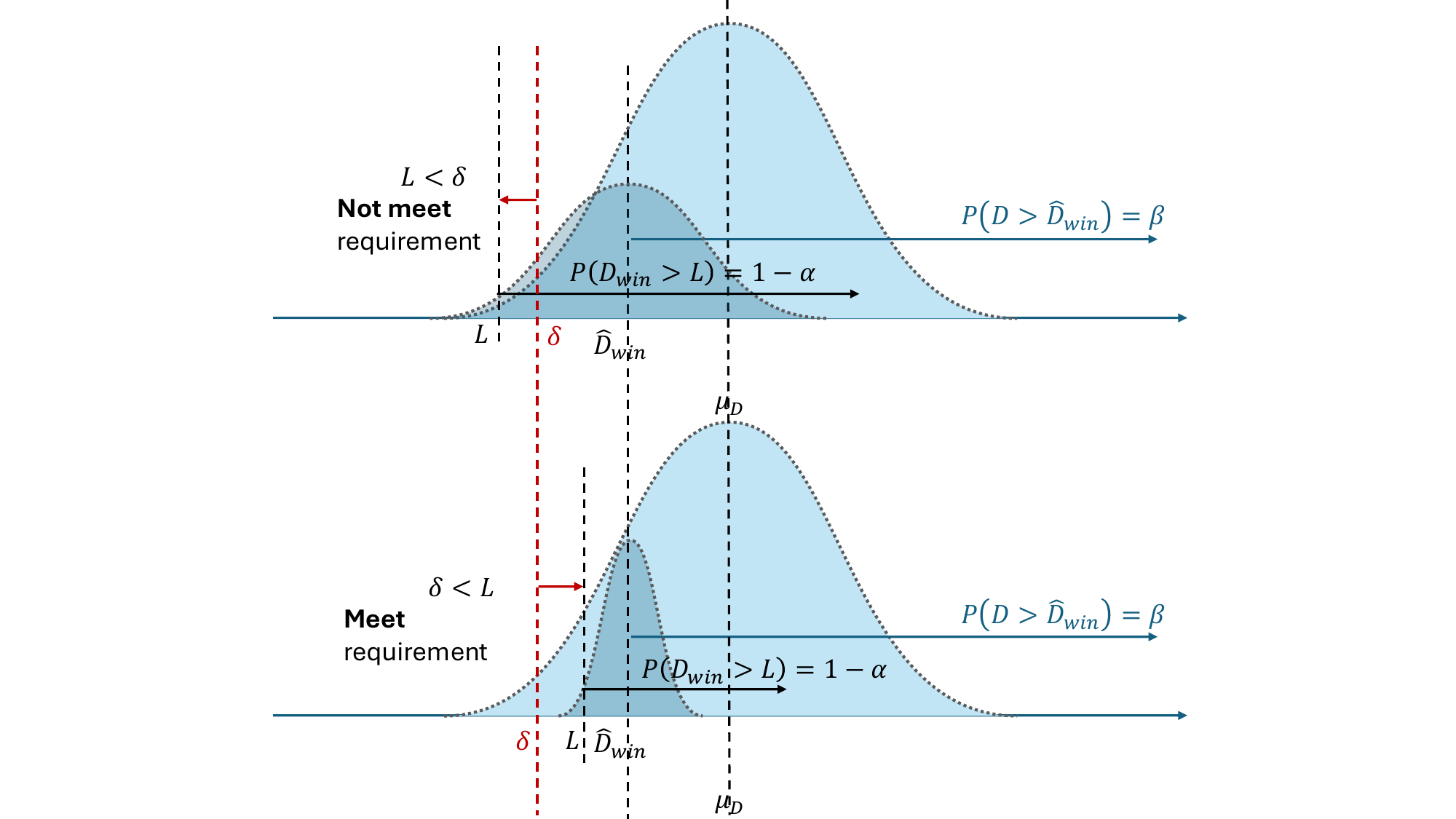}
    \caption[Illustrating for individual benefit analysis framework.]{In reality, Nature does not reveal the true (light blue shaded) density of $D$ to us, so we conduct randomized controlled trials (RCTs) to gain insight into Nature.  Given a desired \emph{Win probability} $\beta$, one can estimate a best-achievable \emph{Win margin} $\hat{D}_{win}$ from each RCT, leading to a (dark blue shaded) density of $\hat{D}_{win}$.  To guarantee that the incorrect decision rate is no more than $\alpha$, we calculate a lower bound $L$ based on $\hat{D}_{win}$ such that, if we decide that the patients' requirement is met only when $L > \delta$, then across infinitely many RCTs the rate of incorrect decision is no more than $\alpha$.  In this figure, only the bottom plot leads us to decide the patients' requirement is met.}
    \label{fig:IndividualBenefitPlot}
\end{figure}

Here is the intuition for the answer to \hyperlink{q:main}{the key question of interest}, illustrated by Figure \ref{fig:IndividualBenefitPlot}.
We take using the individual benefit lower bound to answer the question as an example, and it is similar for using the individual benefit probability to answer the key question.

In the population space, we define the unobserved lower $(1-\beta)$th quantile, denoted by $\hat{D}_{\text{win}}$, of the distribution of the individual benefit contrast $D$. 
$\hat{D}_{win}$ is the best-achievable \emph{Win margin} for \emph{Win probability} $\beta$.
That is, for a randomly selected pair of patients from an infinite population, the probability that their individual benefit contrast exceeds $\hat{D}_{\text{win}}$ is $\beta$, $P(D>\hat{D}_{\text{win}}) = \beta$.

We estimate $\hat{D}_{\text{win}}$ by repeatedly replicating the observed study, each time with $n$ pairs of patients.

The individual benefit lower bound $L$ is defined as the lower $\alpha$-quantile of $\hat{D}_{\text{win}}$, such that the interval $[L, +\infty)$ covers the true value of $\hat{D}_{\text{win}}$ with probability at least $1 - \alpha$.

Taking into account both the randomness in the distribution of $D$ and the sampling variability, the answer to \hyperlink{q:main}{the key question of interest} is ``yes'' if $L > \delta$, and ``no'' if $L < \delta$. 
The probability of incorrectly declaring that the requirement is met is at most $5\%$.

\subsection{Counterfactual individual benefit analysis in ETZ framework}
\label{subsec:counterfactual_individual_benefit}
Section \ref{sec:IBMethodFactual} considers factual efficacy through a random-pair contrast: one future patient receives $Rx$ and an independently selected future patient receives $C$. We now extend the same individual-benefit question to the counterfactual setting, in which the two potential outcomes belong to the same patient.

For patient $i$, let $Y_i^{[change]}(Rx)$ and $Y_i^{[change]}(C)$ denote the potential change-from-baseline outcomes under $Rx$ and $C$, respectively. Define the counterfactual individual efficacy contrast as
\[
D_i^{cf}
=
Y_i^{[change]}(Rx)-Y_i^{[change]}(C).
\]
For a prespecified Win margin $\delta$, the corresponding counterfactual Win probability is
\[
\beta^{cf}(\delta)
=
P\left\{D_i^{cf}>\delta\right\}.
\]
Thus, $\beta^{cf}(\delta)$ is the proportion of patients in the target population whose outcome would be better under $Rx$ than under $C$ by more than $\delta$.

Under ignorable randomization, the marginal means of the two potential outcomes are identified by the corresponding treatment arms. Therefore,
\[
E\left(D_i^{cf}\right)
=
E\left\{Y_i^{[change]}(Rx)\right\}
-
E\left\{Y_i^{[change]}(C)\right\}
=
\mu^{Rx[change]}-\mu^{C[change]}.
\]

Hence, the factual and counterfactual efficacy contrasts have the same population means. However, the counterfactual uncertainty can be reduced significantly through ETZ decomposition.

To see this difference, consider the ETZ representation proposed in \cite{wang2025counterfactual}. 
Before treatment is given,
\[
Y_i^{[1]}
=
Z_i+E_i^{[1]},
\]
and at the milestone visit, under treatment $Trt\in\{Rx,C\}$,
\[
Y_i^{Trt[m]}
=
Z_i+\mathrm{Traj}_i^{Trt[m]}+E_i^{Trt[m]}.
\]
Therefore,
\[
Y_i^{[change]}(Trt)
=
\mathrm{Traj}_i^{Trt[m]}
+
E_i^{Trt[m]}
-
E_i^{[1]}.
\]

For the factual contrast $D$ defined in Section 3.1, the two change-from-baseline outcomes come from different patients, say patient $r$ receiving $Rx$ and patient $s$ receiving $C$. Thus,
\[
D
=
\mathrm{Traj}_r^{Rx[m]}
-
\mathrm{Traj}_s^{C[m]}
+
E_r^{Rx[m]}
-
E_s^{C[m]}
-
E_r^{[1]}
+
E_s^{[1]}.
\]
Following the ETZ working assumptions that measurement error is independent of trajectory, measurement errors are i.i.d. across visits, and the two treatment arms have a common trajectory variance, we obtain
\[
\operatorname{Var}(D)
=
2\operatorname{Var}(\mathrm{Traj})
+
4\operatorname{Var}(E).
\]

For the counterfactual contrast, both potential outcomes refer to the same patient. The same pre-treatment baseline, including the same baseline measurement error, appears in both potential change-from-baseline outcomes and therefore cancels. Consequently,
\[
D_i^{cf}
=
\mathrm{Traj}_i^{Rx[m]}
-
\mathrm{Traj}_i^{C[m]}
+
E_i^{Rx[m]}
-
E_i^{C[m]}.
\]
It follows that
\[
\operatorname{Var}\left(D_i^{cf}\right)
=
2\operatorname{Var}(\mathrm{Traj})
+
2\operatorname{Var}(E)
-
2\operatorname{Cov}\left(
\mathrm{Traj}_i^{Rx[m]},
\mathrm{Traj}_i^{C[m]}
\right).
\]
Therefore,
\[
\operatorname{Var}(D)
-
\operatorname{Var}\left(D_i^{cf}\right)
=
2\operatorname{Var}(E)
+
2\operatorname{Cov}\left(
\mathrm{Traj}_i^{Rx[m]},
\mathrm{Traj}_i^{C[m]}
\right).
\]

This is the uncertainty reduction obtained by moving from the factual random-pair comparison to the counterfactual same-patient comparison. The first term arises because the common baseline measurement cancels. The second term reflects the extent to which the same patient's potential trajectories under $Rx$ and $C$ move together.

A transparent working case is
\[
\operatorname{Cov}\left(
\mathrm{Traj}_i^{Rx[m]},
\mathrm{Traj}_i^{C[m]}
\right)
=
0,
\]
which is plausible when $Rx$ and $C$ act through unrelated mechanisms. Then
\[
\operatorname{Var}\left(D_i^{cf}\right)
=
2\operatorname{Var}(\mathrm{Traj})
+
2\operatorname{Var}(E),
\]
so counterfactual uncertainty is smaller than factual uncertainty by exactly
\[
2\operatorname{Var}(E).
\]
If the two potential trajectories are positively correlated, as may occur when $Rx$ and $C$ have related mechanisms of action, the counterfactual variance is reduced further. Since this cross-treatment trajectory covariance is not directly observed in a parallel-arm factual RCT, the zero-covariance case provides a transparent working assumption and, when non-negative covariance is scientifically plausible, a conservative counterfactual variance calculation.

Under the same Normal working model used in Section 3.1, write
\[
D_i^{cf}
\sim
N\left(
\mu_D,
\sigma_{D,cf}^2
\right),
\]
where
\[
\mu_D
=
\mu^{Rx[change]}-\mu^{C[change]}
\]
and
\[
\sigma_{D,cf}^2
=
\operatorname{Var}\left(D_i^{cf}\right).
\]
Then the counterfactual Win probability is
\[
\beta^{cf}(\delta)
=
P\left(D_i^{cf}>\delta\right)
=
\Phi\left(
\frac{\mu_D-\delta}{\sigma_{D,cf}}
\right).
\]
For a required Win probability $\beta$, the corresponding best-achievable counterfactual Win margin is
\[
D_{win}^{cf}(\beta)
=
\mu_D
-
\Phi^{-1}(\beta)\sigma_{D,cf}.
\]

The confidence-based construction of the individual benefit probability $p_{ib}$ and individual benefit lower bound $L$ in Section 3.1 can therefore be applied to the counterfactual setting by replacing the factual random-pair distribution of $D$ with the counterfactual same-patient distribution of $D_i^{cf}$. We denote the resulting quantities by $p_{ib}^{cf}$ and $L^{cf}$, respectively.

The distinction is important. The factual analysis quantifies the probability that a randomly selected patient receiving $Rx$ outperforms an independently selected patient receiving $C$, whereas the counterfactual analysis quantifies the proportion of patients who would themselves benefit from switching from $C$ to $Rx$ by more than the prespecified margin $\delta$.

\subsection{Example}\label{sec:IBExample}
We consider an Alzheimer’s disease (AD) patient-level dataset from AbbVie as an example. 
The study, described in \cite{schnell2017subgroup}, is a Phase II clinical trial comparing three dose levels of a novel treatment (doses 1, 2, and 3) against a placebo control (dose 0). 
An additional active control arm (dose 4) was also included in the trial. 
The primary outcome measure is the ADAS-Cog11 score, assessed over a study duration of 24 weeks.

To illustrate the analysis, we restrict attention to male patients and fit a mixed model for repeated measures (MMRM) to the longitudinal ADAS-Cog11 outcomes. The model includes fixed effects for treatment, visit, and their interaction, and employs a factor-analytic covariance structure of order 1 (FA(1)) to account for within-subject correlations over time. The resulting least squares mean (LSMeans) estimates for the difference in change from baseline between each treatment arm and the placebo are summarized in Table~\ref{tab:mmrm_lsmeans}.

\begin{table}[htbp]
\centering
\renewcommand{\arraystretch}{1.15}
\setlength{\tabcolsep}{6pt}
\begin{tabular}{lcc}
\toprule
\textbf{Comparison} & \textbf{Estimate} & \textbf{Standard Error} \\
\midrule
Dose 1 vs Dose 0 & -1.7567 & 1.6700 \\
Dose 2 vs Dose 0 & -3.7103 & 1.6851 \\
Dose 3 vs Dose 0 & -3.6046 & 1.9467 \\
Dose 4 vs Dose 0 & -5.6667 & 1.7562 \\
\bottomrule
\end{tabular}
\caption{Difference in change from baseline (LSMeans estimates) from the AbbVie dataset, obtained using a mixed model for repeated measures (MMRM). Dose groups are defined as follows: Dose 0 = placebo; Dose 1 = low-dose test treatment; Dose 2 = medium-dose test treatment; Dose 3 = high-dose test treatment; Dose 4 = active control.}
\label{tab:mmrm_lsmeans}
\end{table}

\begin{table}[htp]
\centering
\begin{minipage}{0.48\textwidth}
\centering
\textbf{Summary statistics}\\[4pt]
\begin{tabular}{lrr}
\toprule
 & Variance & SD \\
\midrule
Visit 1 & 122.684 & 11.076 \\
Visit 2 & 150.002 & 12.248 \\
Change  & 41.127  & 6.413 \\
\bottomrule
\end{tabular}
\end{minipage}
\hspace{0.02\textwidth}
\begin{minipage}{0.48\textwidth}
\centering
\textbf{ETZ variance components}\\[4pt]
\begin{tabular}{lrr}
\toprule
 & Variance & SD \\
\midrule
Intercept Z       & 115.78  & 10.76 \\
Trajectory        & 27.318  & 5.227 \\
Measurement Error & 6.905   & 2.628 \\
\bottomrule
\end{tabular}
\end{minipage}

\caption[Apply ETZ decomposition to endpoint ADASCog11 from AbbVie data.]{Apply ETZ decomposition to endpoint ADASCog11 from AbbVie data. Variances of first visit (baseline), milestone visit and change-from-baseline and variance of Intercept, Trajectory and Measurement Error are reported. There are two visits in total, so Visit 2 is the milestone visit.}
\label{tab:ADASCogAbbVie}
\end{table}

Since higher ADAS-Cog11 scores indicate worse cognitive function, while our method constructs a lower bound, we reverse the sign of the outcome for ease of interpretation. Specifically, we define the transformed outcome as the negative ADAS-Cog11 score, so that larger values correspond to better cognitive status.

Here, we take the comparison between treatment 1 and treatment 0 as an example to compute the individual benefit probability and the individual benefit lower bound.
We set the confidence level \(1-\alpha = 95\%\), win margin \(\delta = 0\) and win probability as \(80\%\).
The \hyperlink{q:main}{key question of interest} becomes
\begin{quote}
\textit{Is it true that at least 80\% of patients will derive a higher benefit from receiving \(Rx\) than \(C\)?}
\end{quote}
and we control the probability of falsely declaring that this requirement is met at 5\%.
To align with the MMRM output, we approximate the effective sample size using \(n = Var(D)/SE^2\).

\paragraph{Factual individual benefit analysis.}
We start from the factual random-pair contrast, in which the patient factually receiving $Rx$ and the patient factually receiving $C$ are independently selected.

Following the method described in Section \ref{sec:IBMethodFactual} and referring to the information summarized in Table \ref{tab:mmrm_lsmeans}, we can compute the individual benefit probability and individual benefit lower bound for each comparison in the AbbVie datasets.
The results are summarized in Table \ref{tab:IB_results}.

Unfortunately, the answers to the question of interest are all ``no'' when $\delta = 0$ and $\beta = 80\%$.
However, we can still obtain useful information from Table~\ref{tab:IB_results}.
For example, when $\delta = 0$ and $\beta = 0.6$, only Dose~4 gives a ``yes'' answer to the key question of interest.
When $\beta = 0.8$ and $\delta = -7$, only Dose~4 gives a ``yes'' answer to the key question of interest.

In this case, the requirement looks a bit ``strict'' and it is hard to get a ``yes'' answer.
This may be because the efficacy in the overall population is driven downward by one particular subgroup. 
Hence, it is worth further investigating the conditional efficacy within subgroups and exploring whether the answer to the key question of interest can be changed to ``yes'' when subgroup-specific effects are considered in future analyses.
\begin{table}[htbp]
\centering
\renewcommand{\arraystretch}{1.12}
\setlength{\tabcolsep}{6pt}
\begin{tabular}{lcccc}
\toprule
\textbf{Comparison} & $\mathbf{n}$ & $\mathbf{p_{ib}}~(\delta = 0)$ & $\mathbf{L}~(\beta = 0.8)$ & Mann-Whitney\\
\midrule
Dose 1 vs Dose 0 & 29.5 & 0.455 & -9.64 & 0.596 \\ 
Dose 2 vs Dose 0 & 29.0 & 0.535 & -7.73 & 0.686 \\ 
Dose 3 vs Dose 0 & 21.7 & 0.510 & -8.59 & 0.713 \\ 
Dose 4 vs Dose 0 & 26.7 & 0.607 & -5.98 & 0.737 \\ 
\bottomrule
\end{tabular}
\caption[Individual benefit metrics under the individual benefit framework.]{Individual benefit metrics under the individual benefit framework in the factual setting.
The effective sample size $n$ is approximated via
$n=\operatorname{Var}(D)/SE^2$,
linking the MMRM-derived standard errors to an equivalent sampling formulation.
For a prespecified Win margin $\delta$, $p_{ib}$ denotes the $100(1-\alpha)\%$
lower confidence bound for the population Win probability
$\Pr(D>\delta)$.
For a prespecified Win probability $\beta$, $L$ denotes the corresponding
$100(1-\alpha)\%$ lower confidence bound for the best-achievable Win margin,
defined through $\Pr(D>D_{\mathrm{win}})=\beta$.
Here, $\delta=0$, $\beta=0.8$, and $1-\alpha=0.95$.
Dose 0 corresponds to placebo; Doses 1--4 represent the treatment arms,
with Dose 4 being the active control.}
\label{tab:IB_results}
\end{table}

\paragraph{Counterfactual individual benefit analysis.}
The analysis above is based on the factual random-pair setting. We now repeat the individual benefit analysis under the counterfactual setting, where the potential outcomes under $Rx$ and $C$ are compared for the same patient.

From the ETZ decomposition reported above,
\[
\operatorname{Var}(\mathrm{Traj})=27.318
\]
and
\[
\operatorname{Var}(E)=6.905.
\]
Therefore, the variance of the factual difference in change-from-baselines is
\[
\begin{aligned}
\operatorname{Var}(D)
&=
2\operatorname{Var}(\mathrm{Traj})
+
4\operatorname{Var}(E)\
&=
2(27.318)+4(6.905)\
&=
82.256.
\end{aligned}
\]
The variance of the counterfactual difference between change-from baselines of same patient under $ Rx $ and $ C $ is
\[
\begin{aligned}
\operatorname{Var}(D^{cf})
&=
2\operatorname{Var}(\mathrm{Traj})
+
2\operatorname{Var}(E)\
&=
2(27.318)+2(6.905)\
&=
68.446.
\end{aligned}
\]

Applying the same $95\%$ confidence construction used for the factual individual benefit analysis, but replacing the factual distribution of $D$ with the counterfactual distribution of $D_i^{cf}$, gives the results in Table~\ref{tab:CounterfactualIBResults}.

\begin{table}[htbp]
\centering
\renewcommand{\arraystretch}{1.12}
\setlength{\tabcolsep}{6pt}
\begin{tabular}{lccc}
\toprule
\textbf{Comparison} 
& $\mathbf{n}$ 
& $\mathbf{p_{ib}^{cf}}~(\delta = 0)$ 
& $\mathbf{L^{cf}}~(\beta = 0.8)$ \\
\midrule
Dose 1 vs Dose 0 & 29.5 & 0.462 & -8.65 \\
Dose 2 vs Dose 0 & 29.0 & 0.549 & -6.73 \\
Dose 3 vs Dose 0 & 21.7 & 0.524 & -7.52 \\
Dose 4 vs Dose 0 & 26.7 & 0.627 & -4.95 \\
\bottomrule
\end{tabular}

\caption[Individual benefit metrics under the counterfactual individual benefit framework.]
{Individual benefit metrics under the individual benefit framework in the counterfactual setting.
The effective sample size $n$ is inherited from the corresponding factual analysis and is approximated via
$n=\operatorname{Var}(D)/SE^2$,
linking the MMRM-derived standard errors to an equivalent sampling formulation.
Under the working assumption that the potential trajectories under $Rx$ and $C$ are uncorrelated,
for a prespecified Win margin $\delta$, $p^{cf}_{ib}$ denotes the $100(1-\alpha)\%$
lower confidence bound for the counterfactual population Win probability
$\Pr(D^{cf}>\delta)$.
For a prespecified Win probability $\beta$, $L^{cf}$ denotes the corresponding
$100(1-\alpha)\%$ lower confidence bound for the counterfactual best-achievable Win margin,
defined through $\Pr(D^{cf}>D^{cf}_{\mathrm{win}})=\beta$.
Here, $\delta=0$, $\beta=0.8$, and $1-\alpha=0.95$.
Dose 0 corresponds to placebo; Doses 1--4 represent the treatment arms,
with Dose 4 being the active control.}
\label{tab:CounterfactualIBResults}
\end{table}

Compared with the factual analysis, both counterfactual individual benefit metrics move in the favorable direction for every treatment comparison: $p_{ib}^{cf}$ increases and $L^{cf}$ becomes larger. 
This occurs even though the estimated mean efficacy is unchanged, because ETZ-based counterfactual uncertainty quantification removes the baseline measurement-error contribution that arises when two different patients are compared.

For the original requirement $\delta=0$ and $\beta=0.8$, however, the answer remains ``no'' for all four dose comparisons. Even after counterfactual uncertainty reduction, none of the $95\%$ lower confidence bounds $p_{ib}^{cf}$ reaches $0.8$, and none of the corresponding $L^{cf}$ values exceeds zero.

The difference between factual and counterfactual inference becomes visible for less stringent requirements. For example, when $\delta=0$ and $\beta=0.6$, Dose 4 satisfies the requirement because
\[
p_{ib}^{cf}=0.627>0.6.
\]
More importantly, consider $\beta=0.8$ and $\delta=-7$. Under the factual analysis, only Dose 4 satisfies the requirement: the factual lower bound for Dose 2 is $-7.73$, which does not exceed $-7$. Under counterfactual uncertainty quantification, however,
\[
L^{cf}({\mathrm{Dose\ 2}})=-6.73>-7
\]
and
\[
L^{cf}({\mathrm{Dose\ 4}})=-4.95>-7.
\]
Therefore, both Dose 2 and Dose 4 satisfy the requirement in the counterfactual setting.

This comparison illustrates the practical consequence of counterfactual uncertainty quantification. The treatment-effect point estimate itself is unchanged, but a reduction in the uncertainty relevant to the same-patient treatment comparison can change whether a clinically specified individual-benefit requirement can be supported with the required confidence.

If the potential trajectories under $Rx$ and $C$, $\operatorname{Cov}
\left(\mathrm{Traj}_i^{Rx[m]},\mathrm{Traj}_i^{C[m]}\right)$, are positively correlated, $\operatorname{Var}(D_i^{cf})$ would be further reduced. Hence, the uncertainty reduction calculated under the zero-covariance assumption is conservative.
 Because this cross-treatment trajectory covariance is not directly observed in a parallel-arm randomized trial, the present example uses zero covariance as the reference counterfactual calculation.
\section{Discussion and future work}
\label{sec:discussion_future_work}

This work develops an ETZ-powered framework for answering a patient-centered personalized medicine question in the overall population: whether at least a prespecified proportion $\beta$ of patients would benefit from receiving $Rx$ rather than $C$ by more than a clinically meaningful margin $\delta$. 

The central conclusion is that personalized medicine cannot be supported by mean-based efficacy comparisons alone. 
Mean-based approaches, including the optimal individualized treatment rule, assign treatments by maximizing conditional expected outcomes \cite{zhao2015new}. However, the counterexamples presented in this paper show that a favorable mean treatment effect can coexist with a much less favorable individual benefit probability. Consequently, a substantial proportion of patients may still fail to benefit from the treatment selected by a mean-based rule. Average superiority therefore does not answer the probability-based question that is directly relevant to individual patients.

To address this limitation, we quantify personalized benefit using the individual benefit probability $p_{ib}$ and the individual benefit lower bound $L$ in both factual and counterfactual settings. These quantities support a confidence-based decision on whether
\[
\Pr(D>\delta)\geq\beta,
\]
with the incorrect-decision probability controlled at a prespecified level. The ETZ decomposition separates trajectory variability from measurement error and, in the same-patient counterfactual estimand, allows the common baseline measurement error to cancel, thereby reducing uncertainty \cite{wang2025counterfactual}.

Conformal inference instead constructs intervals for a future patient's outcome or individual treatment effect \cite{lei2021conformal}, whereas our framework targets the proportion of the overall population whose individual benefit exceeds a clinically meaningful margin. By adopting a Normal working model and using ETZ-powered counterfactual uncertainty quantification, our method can provide narrower confidence bounds when these assumptions are adequate.

The present paper focuses on personalized benefit in the overall population and does not yet address the subgroup setting that is common in personalized medicine. Future research related to biomarker-defined subgroups should first follow the \emph{equipoise principle}, under which efficacy is assessed and reported for the biomarker-positive subgroup, the biomarker-negative subgroup, and the overall population \cite{Han2020ExactSimultaneousCI}. This avoids allowing a favorable result in one subgroup to drive the overall conclusion while obscuring a lack of benefit, or possible harm, in the complementary subgroup.

Subgroup analysis research should also follow the \emph{logic-respecting principle} \cite{liu2022rejoinder}. In particular, the overall efficacy assessment should remain within the range of the corresponding subgroup-specific assessments; collapsibility alone is not sufficient to guarantee this logical relationship \cite{colnet2024riskratiooddsratio}. Future work should therefore develop subgroup-specific individual benefit probabilities and confidence bounds using logic-respecting efficacy measures and inferential procedures. Combining these principles with ETZ-powered uncertainty decomposition would extend the current overall-population framework to principled subgroup targeting while preserving coherent, patient-centered treatment decisions.

\section*{Acknowledgments}

The ETZ modeling thinking started with Haiyan Xu saying “Dr. Hsu, you only need three numbers!”  

\section*{Declaration of AI Use}

ChatGPT (OpenAI) was used solely for language editing during the preparation of this manuscript. All revisions were reviewed and approved by the authors.





\bibliographystyle{Chicago}

\bibliography{Xreference}
\appendix
\section{Features of online testing}
\label{app:OnlineTesting}
Online testing, particularly A/B testing, was started by \cite{kohavi2020trustworthy}. 
Each A/B test, comparing the \(n^{th}\) pair of landing pages \(A_n/B_n\), is a Randomized Controlled Trial (RCT) in which subjects are online visitors to a website or potential customers to a service.  
Companies that conduct real online testing include tech firms like Airbnb, Amazon, Facebook, Google, Lyft, LinkedIn,  Microsoft, Netflix, Spotify, Twitter/X, and Uber.
Amazon, for example, runs thousands of internal experiments across its retail site, comparing pairs of web pages with slightly different layouts and button placements.
In addition, Amazon provides an A/B testing tool called \textit{Manage Your Experiments} for their sellers.  
This system \textit{randomly} splits traffic 50/50 to compare two versions of listing elements (product images, titles, bullet points, descriptions, and pricing).  
Key comparison metrics are 
\begin{enumerate}
    \item[CTR] \textit{Click Through Rate}: how often shoppers click a listing after seeing it in search results
    \item[ATC] \textit{Add-to-Cart Rate}: whether the listing persuades shoppers to begin the buying process
    \item[CVR] \textit{Conversion Rate}: how many shoppers who view a listing end up purchasing
\end{enumerate}

To us, two key characteristics of real online testing are 
\begin{description}
    \item \textit{No going back}. Keep the decisions that have been made (e.g., those regarding button placement).  
    \item \textit{Going forward}. Use knowledge from past experiments to design and analyze future experiments.  
\end{description}

\section{A counterexample for individualized treatment rules at the average depth}\label{sec:CounterexampleAvg}

Consider the no-covariate setting summarized in Table \ref{tab:unequal_sd_counterexample}. Since
\[
E\!\left[Y_i(Rx)\right] = 2 > 1 = E\!\left[Y_i(C)\right],
\]
the OITR assigns $Rx$ to every patient at the average depth.

\begin{table}[htbp]
\centering
\begin{tabular}{|c|c|c|c|c|c|}
\hline
meanC & meanRx & sdC & sdRx & $\delta$ & percentage \\ \hline
1 & 2 & 1.5 & 3 & 0 & 61.72\% \\ \hline
\end{tabular}
\caption{A counterexample at the average depth. The percentage here denotes $P\{Y_i(Rx)-Y_i(C)>\delta\}$.}
\label{tab:unequal_sd_counterexample}
\end{table}

Assume
\[
Y_i(Rx)\sim N(2,9),
\qquad
Y_i(C)\sim N(1,2.25),
\]
and, for simplicity, assume that $Y_i(Rx)$ and $Y_i(C)$ are independent. Then
\[
\Delta_i = Y_i(Rx)-Y_i(C)\sim N(1,11.25).
\]
Therefore,
\[
P(\Delta_i>0)
=
P\!\left(Z>\frac{0-1}{\sqrt{11.25}}\right)
=
P\!\left(Z>-\frac{1}{\sqrt{11.25}}\right)
=
\Phi\!\left(\frac{1}{\sqrt{11.25}}\right)
=
\Phi(0.2981)
\approx 0.6172,
\]
where $Z\sim N(0,1)$ and $\Phi(\cdot)$ denotes the standard normal cumulative distribution function. Hence, although $Rx$ is preferred at the average depth, the proportion of patients who actually benefit from $Rx$ is only $61.72\%.$ Equivalently, about
\[
1-0.6172 = 0.3828 = 38.28\%
\]
of patients do not benefit from $Rx$ at the individual level.
\end{document}